\documentclass[12pt,twocolumn]{iopart}
\usepackage{graphicx}

\begin{document}

\title[Fluctuations of PC]{Fluctuations of persistent current}

\author{A. G. Semenov$^1$
and A. D. Zaikin $^{1,2}$
}

\address{
  $^1$ I.E.Tamm Department of Theoretical Physics, P.N.Lebedev
Physics Institute, 119991 Moscow, Russia}
\address{
  $^2$  Institute for Nanotechnology, Karlsruhe Institute of
Technology (KIT), 76021 Karlsruhe, Germany
}

\ead{semenov@lpi.ru}


\submitto{JPCM}

\begin{abstract}

We theoretically analyze equilibrium fluctuations of
persistent current (PC) in nanorings. We demonstrate that these
fluctuations persist down to zero temperature provided the current
operator does not commute with the total Hamiltonian of the
system. For a model of a quantum particle on a ring we explicitly
evaluate PC noise power which has the form of sharp peaks at
frequencies set by the corresponding interlevel distances.
In
rings with many conducting channels a much smoother and broader
PC noise spectrum is expected. A specific feature of PC noise is that
its spectrum can be tuned by an external magnetic flux
indicating the presence of quantum coherence in the system.
\end{abstract}

\maketitle

\section{Introduction}

Meso- and nanorings formed by normal conductors and pierced by
external magnetic flux develop persistent currents \cite{Imry}.
This fundamentally important equilibrium effect is a direct
consequence of quantum coherence of electrons which -- at low
enough temperatures -- may persist up to distances exceeding the
perimeter of such rings.

Does persistent current (PC) fluctuate? At the first sight it
might appear reasonable to assume that at least for $T \to 0$ no
such fluctuations could occur. Indeed, while at non-zero $T$
thermal fluctuations of PC should be present \cite{Moskalets},
in the zero temperature limit the system approaches its (non-degenerate)
ground state and, hence, no PC fluctuations would be possible.

Below we will demonstrate that in many cases it is not so. Namely,
no PC fluctuations are expected in the zero temperature limit only
provided the current operator commutes with the total Hamiltonian
of the system, otherwise fluctuations of persistent current can occur even in the ground state exactly at $T=0$. Theoretical and experimental investigations of such PC fluctuations can give important additional information about the ground state properties of meso- and nanorings not contained in the average value of PC.

Note that fluctuations of PC in the ground state may be induced
provided the ring interacts with some quantum dissipative
environment. In this case such interaction is responsible for (i)
phase-breaking effects implying suppression of both quantum
coherence and PC and (ii) non-vanishing fluctuations of PC down to $T \to 0$. E.g. it was demonstrated \cite{Buttiker} that
interaction with Caldeira-Leggett environment decreases the
average value of PC and simultaneously {\it increases} PC
fluctuations which are directly related to fluctuations in the
environment itself. Such effects are of importance, e.g., for PC qubits which quantum states can be entangled with those of environment \cite{Jordan}. Fluctuations of PC down to $T \to 0$ could also occur provided the number
of particles in a ring fluctuates due to its interaction
with some reservoir \cite{CB}.

The situation considered here is entirely different: We do not
assume the presence of interaction or particle exchange with any environment at all.
Accordingly, quantum coherence of the system is fully preserved
and no PC suppression takes place. As it will be demonstrated
below, quantum coherence implies the possibility of tuning of
PC fluctuations by an external magnetic flux applied to the system.

\section{The model and general relations}

For definiteness, let us
consider a simple model of a quantum particle with mass $M$ on a 1d
ring of radius $R$ pierced by magnetic flux $\Phi$, see, e.g.,
Fig. 1. The particle position on the ring is parametrized by the
angle $\theta$ which will be the quantum mechanical variable of
interest in our problem. The Hamiltonian for this system reads
\begin{equation}
  \hat H=\frac{(\hat \phi +\phi_x)^2}{2MR^2}+U(\theta ),
\end{equation}
\begin{figure}
\begin{center}
\includegraphics[width=3.375in]{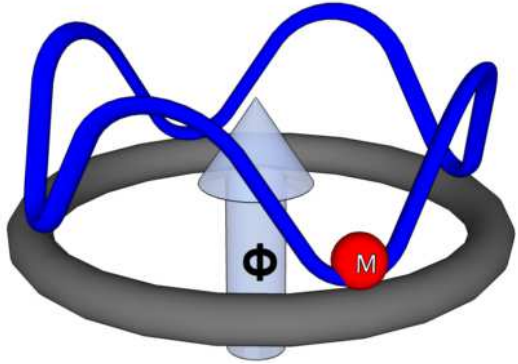}
\end{center}
\label{f1}
\caption{The system under consideration:
A particle on a ring in the presence of a periodic potential. The
ring is pierced by the magnetic flux.}
\end{figure}
where $\hat \phi =-i\frac{\partial}{\partial\theta}$ is the
(dimensionless) flux operator, $U(\theta)$ defines the potential
profile for our particle, $\phi_x=\Phi/\Phi_0$ and $\Phi_0$ is the
flux quantum. Within this model PC was previously studied in a
number of papers \cite{GZ98,Paco,GHZ,HlD,KH,SZ} in the presence of
various dissipative environments. In addition, the model discussed
here could be of interest for the problem of PC in superconducting
nanorings in the presence of quantum phase slips \cite{AGZ}.

For our problem the current operator in the standard Schrodinger
representation is defined as
\begin{equation}
    \hat I=\frac{e}{2\pi}\dot{\hat \theta}=\frac{ie}{2\pi}[\hat H,\hat\theta ]=\frac{e(\hat\phi +\phi_x)}{2\pi MR^2}.
\end{equation}
Switching to the Matsubara representation
\begin{equation}
   \hat I_M(\tau)=e^{\tau\hat H}\hat I e^{-\tau\hat H},
\end{equation}
we define the current-current correlation function
\begin{eqnarray}
  \Pi (\tau)=\langle\langle\mathcal T\hat I_M(\tau)\hat
  I_M(0)\rangle\rangle=T\sum\limits_{n=-\infty}^\infty\Pi_{i\omega_n}e^{-i\omega_n\tau},
\label{corrM}
\end{eqnarray}
which describes equilibrium current noise. Here $\mathcal T$ is
the time-ordering operator, $\omega_n=2\pi nT$ are Matsubara
frequencies, $\langle ... \rangle \equiv {\rm tr}(\hat\rho ...)$
denotes averaging with the equilibrium density matrix
$\hat\rho=e^{-\beta\hat H}/\mathcal Z$, where $\mathcal Z={\rm tr}
e^{-\beta\hat H}$ is the grand partition function and $\beta
=1/T$. The symbol $\langle\langle\dots\rangle\rangle$ stands for
irreducible correlators (cumulants), e.g., $\langle \langle\hat
I(\tau )\hat I(0)\rangle\rangle=\langle\hat I(\tau )\hat
I(0)\rangle
 -\langle\hat I(\tau )\rangle\langle \hat I(0)\rangle$ .

Employing the full set of eigenstates $\hat
H|m\rangle=\varepsilon_m(\phi_x)|m\rangle$ after a straightforward
calculation we obtain
\begin{equation}
  \Pi(\tau)=\mathcal P+\tilde \Pi(\tau),
\label{pi}
\end{equation}
where $\mathcal P$ does not depend on imaginary time and reads
\begin{equation}
\mathcal P=\frac{1}{\mathcal Z}\sum\limits_{m}|\langle m|\hat I|m
\rangle|^2 e^{-\beta\varepsilon_m}-\langle\hat I\rangle^2,
\label{Pdef}
\end{equation}
while the Fourier components of $\Pi(\tau)$ are defined as
\begin{equation}
\tilde \Pi_{i\omega_k}=\frac{1}{\mathcal Z}\sum\limits_{m\neq
n}|\langle m|\hat I|n
\rangle|^2\frac{e^{-\beta\varepsilon_n}-e^{-\beta\varepsilon_m}}{i\omega_k+\varepsilon_m-\varepsilon_n}.
\label{cspe1}
\end{equation}

In order to establish the relation between the correlator
(\ref{corrM}) and the current noise power we also define the
Heisenberg operators $\hat I(t)=e^{it\hat H}\hat I e^{-it\hat H}$
and the Keldysh Green function
\begin{equation}
S(t)=\langle\langle\hat I(t)\hat I(0)+\hat I(0)\hat I(t)
\rangle\rangle=\int\frac{d\omega}{2\pi}S_\omega e^{-i\omega t}.
\label{corrK}
\end{equation}
As before, decomposing the result for $S(t)$ into time-independent
and time-dependent contributions we find
\begin{equation}
  S(t)=2\mathcal P+\tilde S(t),
\label{s}
\end{equation}
where $\mathcal P$ is again defined in Eq. (\ref{Pdef}) and the
Fourier components of $\tilde S(t)$ take the form
\begin{eqnarray} \tilde S_\omega=\frac{2\pi}{\mathcal
Z}\sum\limits_{m\neq n}|\langle m|\hat I|n
\rangle|^2\nonumber\qquad\qquad\ \\
\times\left(e^{-\beta\varepsilon_m}+e^{-\beta\varepsilon_n}\right)\delta(\omega+\varepsilon_m-\varepsilon_n).
\label{cspe2}
\end{eqnarray}
Comparing now Eqs. (\ref{cspe1}) and (\ref{cspe2}) we arrive at the
fluctuation-dissipation relation
\begin{equation}
 \tilde S_\omega=2\coth\frac{\beta\omega}{2}\Im \tilde\Pi_{\omega+i0},
\end{equation}
which allows to immediately recover the current noise power from
the correlator (\ref{pi})-(\ref{cspe1}). Note that this relation
links together the quantities $\tilde\Pi$  and $\tilde S$ which --
according to Eqs. (\ref{pi}) and (\ref{s}) -- differ from the the
correlators $\Pi$ and $S$ by the constant in time terms,
respectively $\mathcal P$ and $2\mathcal P$, which produce
singularities in the frequency domain.

We would like to point out that our formalism also allows to
analyze the linear current response to the time-dependent flux
inside the ring and to formally define the ac conductance of the system.
According to the Kubo formula this ac conductance is expressed in terms
of the commutator of the current operators, unlike the noise spectrum
defined by the anticommutator of these operators
(\ref{corrK}). Below we restrict
our attention only to time-independent values $\phi_x$ and do not address
the behavior of the ac conductance which we are not interested in here.

The above exact relations fully determine PC correlators in terms
of the system eigenstates. These relations allow to observe that
as long as the current operator $\hat I$ commutes with the
Hamiltonian of the system the Fourier components (\ref{cspe1}) and
(\ref{cspe2}) vanish identically together with the matrix elements
$\langle m|\hat I|n \rangle$ with $m \neq n$, while the
time-independent term $\mathcal P$ (\ref{Pdef}) tends to zero only
in the zero temperature limit. Thus, in this case no PC
fluctuations can occur at $T \to 0$ and at non-zero temperatures
PC noise does not vanish only in the zero frequency limit,
$S_\omega =2 \mathcal P\delta (\omega)$.

If, however, the operators $\hat I$ and $\hat H$ do not commute
with each other the situation becomes entirely different. In that
case the matrix elements $\langle m|\hat I|n \rangle$ in general
remain non-zero for any pair of eigenstates and, hence, PC
fluctuations may persist down to $T=0$.

For a simple model of Fig. 1 PC correlators can be evaluated
directly from Eqs. (\ref{corrM})-(\ref{cspe2}). In more
complicated situations, however, the above general relations
employing the representation of eigenstates could become less
convenient for practical calculations. For this reason below we
will develop alternative approaches which can also be useful for
the analysis of PC fluctuations.

\section{Free energy and current noise}

It turns out that in both
limits of zero Matsubara frequency and zero imaginary time the
current-current correlator can be conveniently related to the free
energy of the system $\mathcal F=-T\ln \mathcal Z$. Making use of
the expression for PC
\begin{equation}
\langle\hat I\rangle=\frac{e}{2\pi} \frac{\partial\mathcal
F}{\partial\phi_x}
\end{equation}
together with the identity
\begin{equation}
 \frac{\partial (e^{-\beta\hat H})}{\partial\phi_x}=-\int\limits_0^\beta d\tau e^{-(\beta-\tau)\hat H}
 \frac{\partial \hat H}{\partial\phi_x}e^{-\tau\hat H},
\end{equation}
for the second derivative of the free energy with respect to the
flux we obtain
\begin{equation}
 \frac{e^2}{4\pi^2} \frac{\partial^2\mathcal F}{\partial\phi_x^2}=\frac{e^2}{4\pi MR^2}-\int\limits_0^\beta d\tau \frac{{\rm tr}(\hat I
 e^{-(\beta-\tau)\hat H}\hat I e^{-\tau\hat H})}{{\rm tr}(e^{-\beta\hat H})}+\beta\langle\hat
 I\rangle^2.
\end{equation}
From this equation one readily finds
\begin{equation}
 \int\limits_0^\beta d\tau \Pi (\tau ) \equiv \Pi_0=\frac{e^2}{4\pi^2 MR^2}-\frac{e^2}{4\pi^2} \frac{\partial^2\mathcal F}{\partial\phi_x^2}.
\label{Pi0}
\end{equation}

On the other hand, employing the identity
\begin{equation}
\frac{1}{R^2}\frac{\partial}{\partial M}{\rm
tr}\left(e^{-\beta\hat H}\right)=\frac{2\pi^2}{e^2T}{\rm
tr}\left(\hat I^2e^{-\beta\hat H}\right),
\end{equation}
we get
\begin{equation}
\langle\hat I^2\rangle=-\frac{e^2}{2\pi^2
R^2}\frac{\partial\mathcal F}{\partial M}
\end{equation}
and, hence,
\begin{equation}
\Pi (0)=-\frac{e^2}{2\pi^2 R^2}\frac{\partial\mathcal F}{\partial
M}-\frac{e^2}{4\pi^2}\left(\frac{\partial\mathcal
F}{\partial\phi_x}\right)^2. \label{PiM0}
\end{equation}

As in many cases the free energy of the system can be readily
evaluated, Eqs. (\ref{Pi0}) and (\ref{PiM0}) provide a great deal
of information about PC noise. Further simplifications may appear
in the zero temperature limit since in this case the free energy
reduces to the ground state energy $\mathcal F (T \to 0) =
\varepsilon_0 (\phi_x) $. E.g. for a free particle on a ring (i.e.
for $U (\theta )=0$) one has $\varepsilon_0 (\phi_x) =
\phi^2/(2MR^2)$ and, hence, in this case in the limit $T \to 0$
from Eqs. (\ref{Pi0}) and (\ref{PiM0}) one trivially finds
\begin{equation}
\Pi_0=\Pi (0)=0.
\end{equation}
In agreement with our general analysis in the absence of an
external potential and at $T=0$ the correlator (\ref{corrM})
 vanishes identically for all values of $\tau$ implying that no
fluctuations of PC occur in this case. This is because for $U
(\theta )=0$ the current operator commutes with the Hamiltonian.

At non-zero external potentials $U (\theta )\neq 0$, however,
these two operators do not commute anymore and, hence,
fluctuations of PC in general {\it do not vanish} even at very low
$T$. This conclusion can be reached, e.g., from Eq. (\ref{Pi0})
without any additional calculation. Indeed, for $U (\theta )\neq
0$ the ground state energy $\varepsilon_0 (\phi_x)$ deviates from
$ \phi^2/(2MR^2)$ and, hence, $\Pi_0 \neq 0$ down to $T =0$.

\section{Generating functional}

 Let us now formulate a general
technique that will allow to fully describe current fluctuations
of PC within our model. For this purpose we define the generating
functional
\begin{equation}
\mathcal Z[\eta ]=\int\mathcal D\phi\mathcal D\theta
e^{\int\limits_0^\beta d\tau\left(i \phi \dot\theta -\frac{(\phi
+\phi_x)^2}{2MR^2}-U(\theta)-\eta \phi \right)},
\end{equation}
where $\eta (\tau )$ is the source field for the flux variable
$\phi$. Performing integration over $\phi$ we obtain
\begin{equation}
\mathcal Z[\eta ] \sim\int\mathcal D\theta e^{-\int\limits_0^\beta
d\tau\left(\frac{MR^2(\dot \theta+i\eta )^2}{2}+ i\phi_x(\dot
\theta+i\eta )+U(\theta)\right)}.
\end{equation}
Taking the variational derivative of $\mathcal F[\eta
]=-T\ln\mathcal Z[\eta ]$ over the source field $\eta (\tau )$ and
setting this field equal to zero afterwards, we derive the
relation between the expectation values for the current and the
particle ''velocity'' $\dot \theta$:
\begin{equation}
 \langle I(\tau)
 \rangle=\frac{ie}{2\pi}\langle\dot\theta(\tau)\rangle .
\end{equation}
Similarly, the second derivative of $\mathcal F[\eta ]$ with
respect to $\eta (\tau )$ yields the second current cumulant:
\begin{eqnarray}
 \langle\langle I(\tau_1)I(\tau_2) \rangle\rangle=\frac{e^2}{4\pi^2 MR^2}\delta(\tau_1-\tau_2)
 \nonumber\qquad\qquad\\-\frac{e^2}{4\pi^2}\langle\langle\dot\theta(\tau_1)\dot\theta(\tau_2)\rangle\rangle
 .\label{II}
\end{eqnarray}
Analogously one can establish the relations between higher current
and velocity cumulants. Up to some unimportant $\delta$-functions
at coinciding times (which cancel out in the final result as we
will see below) the latter cumulants, in turn, are evaluated from
the relation
\begin{eqnarray} \langle\langle
\dot\theta(\tau_1)...\dot\theta(\tau_N)
\rangle\rangle=(-i)^N\frac{\delta^N\ln\mathcal Z[\zeta ]}{\delta
\zeta (\tau_1)...\delta \zeta (\tau_N)}|_{\zeta =0},
\label{cumulant}
\end{eqnarray}
where $\mathcal Z[\zeta ]$ is the generating functional
\begin{eqnarray}
\mathcal Z[\zeta ]&=&\int_0^{2\pi}
d\theta_0\sum\limits_{m=-\infty}^\infty e^{2\pi im\phi_x}\\
\nonumber &&\times\int\limits^{\theta_0+2\pi m}_{\theta_0}\mathcal
D\theta e^{-\int\limits_0^\beta d\tau\left(\frac{MR^2\dot
\theta^2}{2}+U(\theta)-i\zeta\dot \theta\right)}. \label{genf}
\end{eqnarray}

The above general expressions allow for straightforward evaluation
of all current cumulants thus establishing ''full-counting
statistics'' of PC in our problem.

\section { Current-current correlator}

 Below we will focus our
attention on the current-current correlation function
(\ref{corrM}) which will be evaluated in the specific limiting
case \cite{SZ}
\begin{equation}
U(\theta)=U_0(1-\cos(\kappa\theta)), \quad U_0\gg \kappa^2/(MR^2).
\label{pot}
\end{equation}
In other words, we will assume that the particle confined to the
1D ring is moving in a periodic potential with the distance $2\pi
/\kappa$ between adjacent minima. For $\kappa =1$ our model reduces to that derived for ultra-thin superconducting rings in the presence of quantum phase slips \cite{AGZ}. As indicated in Eq. (\ref{pot}) the potential barriers between these minima are high, in which case the particle moves around the ring due to hopping from one minimum to another. Semiclassically, these hops are described by multi-instanton trajectories \cite{SZ}
\begin{equation}
   \Theta(\tau)=\sum\limits_j\nu_j\tilde\theta(\tau-\tau_j), \quad
   \nu_j=\pm 1,
\label{minst}
\end{equation}
which dominate the path integral (\ref{genf}). Here
$\tilde\theta(\tau)=4\arctan(e^{\Omega\tau})/\kappa$ is
well known kink solution, describing the particle tunneling with
the amplitude
\begin{equation}
\Delta /2=4(\Omega U_0/\pi )^{1/2}e^{-\frac{8U_0}{\Omega}},
\end{equation}
where $\Omega =\kappa \sqrt{U_0/(MR^2)}$. Substituting the
trajectories (\ref{minst}) into  (\ref{genf}) and performing Gaussian integration we get
\begin{eqnarray}
\mathcal Z[\zeta
]=\kappa\sum\limits_{n=0}^{\infty}\sum\limits_{\nu_1=\pm1}..\sum\limits_{\nu_n=\pm
1} \left(\frac{\Delta}{2}\right)^n \int\limits_0^\beta d\tau_1\int\limits_{\tau_1}^\beta
d\tau_2...\int\limits_{\tau_{n-1}}^{\beta}d\tau_n\sum\limits_{m=-\infty}^{\infty}e^{2\pi
im\phi_x} \nonumber\\ \qquad\qquad\qquad\qquad\times
e^{i\sum\limits_j\nu_j\int\limits_0^\beta \zeta (\tau)\dot
{\tilde\theta}(\tau-\tau_j)d\tau }\mathcal
Z_n[\zeta]\delta_{\sum\limits_i \nu_i,m\kappa},
\end{eqnarray}
where the terms
\begin{equation}
 \mathcal Z_n[\zeta]=e^{-\frac12\int \zeta(\tau)G(\tau,\tau')\zeta(\tau')d\tau d\tau'}
\label{Zn}
\end{equation}
are set by Gaussian fluctuations around $n$-instanton trajectories
$\Theta(\tau)$ (\ref{minst}). The correlator
$G(\tau,\tau')=\langle\delta \dot \theta(\tau)\delta \dot
\theta(\tau') \rangle$ can easily be evaluated for a dilute
instanton gas provided both times $\tau$ and $\tau'$ are outside
the instanton cores, i.e.
$|\tau-\tau_j|,|\tau'-\tau_j|\gg\Omega^{-1}$ for every $j$.
In this case $\mathcal Z_n[\zeta ]$ reduces to the generating
functional for a harmonic oscillator $\mathcal Z_0[\zeta ]$
defined by Eq. (\ref{Zn}) with
\begin{equation}
  G(\tau,\tau')\approx \frac{\Omega^3}{2\kappa^2 U_0}
  e^{-\Omega|\tau-\tau'|}+\frac{1}{MR^2}\delta(\tau-\tau'),
\label{gtau}
\end{equation}
where the last expression remains valid for $\beta\Omega\gg 1$ and
$|\tau-\tau'|\ll\beta$. Proceeding analogously to Ref. \cite{SZ}
and employing the Poisson's resummation formula we obtain
 \begin{equation}
 \mathcal Z[\zeta ]=\mathcal Z_{0}[\zeta]\sum\limits_{k=1}^\kappa
 e^{\Delta\int\limits_0^\beta d\tau\cos\left(\frac{2\pi(\phi_x-k)}{\kappa}
 +\int\limits_0^\beta \zeta
 (\tau_1)\dot{\tilde\theta}(\tau-\tau_1)d\tau_1\right)}.
 \label{genfun}
 \end{equation}
In the limit $\zeta \to 0$ Eq. (\ref{genfun}) reduces to the
partition function \cite{SZ} and the average value of PC is
obtained from
 Eqs. (\ref{genfun}) and (\ref{cumulant}) with $N=1$:
 \begin{equation}
I =\frac{e\Delta}{\kappa}
\frac{\sum_{k=1}^{\kappa}\sin\left(\frac{2\pi(\phi_x-k)}{\kappa}\right)
e^{\beta
\Delta\cos\left(\frac{2\pi(\phi_x-k)}{\kappa}\right)}}{\sum_{k=1}^{\kappa}
e^{\beta \Delta\cos\left(\frac{2\pi(\phi_x-k)}{\kappa}\right)}}.
\label{genexp}
\end{equation}
At low temperatures $T \ll \Delta /\kappa^2$ Eq. (\ref{genexp})
reduces to a simple formula
\begin{equation}
I=\frac{e\Delta}{\kappa}\sin\left(\frac{2\pi\phi_x}{\kappa}\right),
\quad -1/2 < \phi_x \leq 1/2, \label{lowT}
\end{equation}
\begin{figure}
\begin{center}
 \includegraphics[width=3.375in]{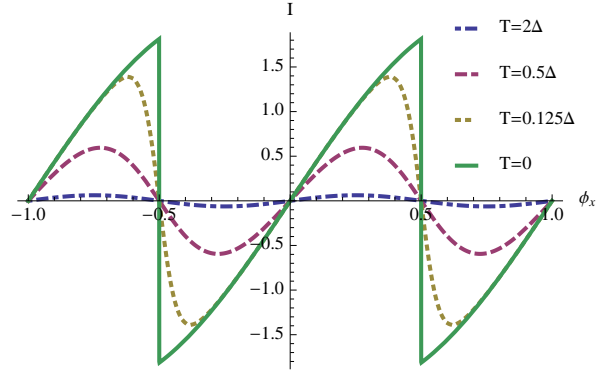}
\end{center}
\label{f3}
\caption{Persistent current $I$ (measured in units $e\Delta /2\pi$) as a function of the magnetic flux $\phi_x$ for $\kappa=3$ at different temperatures:
$T=0$, $T=0.125\Delta$, $T=0.5\Delta$ and $T=2\Delta$.}
\end{figure}
i.e. at $T=0$ the magnitude of PC is
proportional to $\Delta$ while its flux dependence deviates from
a simple sinusoidal form for all $\kappa >1$ and tends to the saw-tooth one in the limit of large $\kappa$. With increasing $T$ PC amplitude decreases and the current-flux dependence gradually approaches the function $I(\phi_x) \propto \sin (2\pi \phi_x)$. These dependencies are
depicted in Fig. 2.

The same equations for $N=2$ yield the second current cumulant
 \begin{equation}
 \Pi (\tau)=\mathcal P+\mathcal P_{osc}(\tau)+\mathcal P_{osc}(\beta-\tau)+\tilde{ \mathcal P}(f(\tau)+
 f(\beta-\tau)),
 \label{mcor}
 \end{equation}
 where for $\Omega^{-1}\leq\tau\leq\beta-\Omega^{-1}$ we find
 \begin{eqnarray}
\mathcal P= \frac{e^2\Delta^2}{\kappa^2}\frac{\sum\limits_{k=1}^\kappa\sin^2\left(\frac{2\pi(\phi_x-k)}{\kappa}\right)
e^{\beta\Delta\cos\left(\frac{2\pi(\phi_x-k)}{\kappa}\right)}}{\sum\limits_{k=1}^\kappa
e^{\beta\Delta\cos\left(\frac{2\pi(\phi_x-k)}{\kappa}\right)}}\nonumber\\
-\frac{e^2\Delta^2}{\kappa^2}\left(\frac{\sum\limits_{k=1}^\kappa\sin\left(\frac{2\pi(\phi_x-k)}{\kappa}\right)
e^{\beta\Delta\cos\left(\frac{2\pi(\phi_x-k)}{\kappa}\right)}}{\sum\limits_{k=1}^\kappa
e^{\beta\Delta\cos\left(\frac{2\pi(\phi_x-k)}{\kappa}\right)}}\right)^2
\label{pppp}
\end{eqnarray} and
\begin{equation}
\tilde{\mathcal
P}=-\frac{e^2\Delta}{\kappa^2}\frac{\sum\limits_{k=1}^\kappa\cos\left(\frac{2\pi(\phi_x-k)}{\kappa}\right)
e^{\beta\Delta\cos\left(\frac{2\pi(\phi_x-k)}{\kappa}\right)}}{\sum\limits_{k=1}^\kappa
e^{\beta\Delta\cos\left(\frac{2\pi(\phi_x-k)}{\kappa}\right)}}.
\label{tpp}
\end{equation}
In Eq. (\ref{mcor}) for $\tau\gg\Omega^{-1}$ we also defined
\begin{eqnarray}
f(\tau)=\frac{\kappa^2}{4\pi^2}\int\limits_{-\infty}^\infty\dot{\tilde
\theta}(\tau_1-\tau)\dot{\tilde \theta}(\tau_1)d\tau_1\simeq
\frac{4\Omega^2\tau}{\pi^2}e^{-\Omega\tau}
\label{ftau}
\end{eqnarray} and
\begin{equation} \mathcal
P_{osc}(\tau)=\frac{e^2\Omega^3}{8\pi^2\kappa^2 U_0}
e^{-\Omega\tau}.
\label{Posc}
\end{equation}
As one could expect from our general analysis in terms of the
exact eigenstates the result (\ref{mcor}) indeed consists of two
different -- time-independent and time-dependent -- contributions.
The meaning of each of these terms can be identified with the aid
of Eqs. (\ref{pi})-(\ref{cspe1}). As we already discussed, exactly
at $T=0$ the time-independent part $\mathcal P$ should vanish,
$\mathcal P=0$. This fact is indeed directly observed from our
result (\ref{pppp}) in the limit $T \to 0$.

We also note that the expression for $\mathcal P$ (\ref{pppp}) can
be established within the effective tight-binding model in which
case the particle successively hops between $\kappa$ nods on the
ring. This observation demonstrates that the term $\mathcal P$ is
universal meaning that it depends only on the tunneling amplitude
$\Delta$ but not on the profile of the periodic potential. Within
the tight-binding model the current operator commutes with the
total Hamiltonian, the current-current correlator does not depend
on $\tau$ and vanishes in the limit $T\to 0$ in accordance with
our general considerations.

At non-zero temperatures, however, the term $\mathcal P$ does not
vanish. At $\Delta /\kappa \ll T \ll \Omega$ from Eq. (\ref{pppp})
we get
\begin{eqnarray}
\mathcal P=  \frac{e^2\Delta^2}{2\kappa}\left(I_0(\beta\Delta)-
I_2(\beta\Delta)\right)\qquad\qquad\nonumber\\-
\frac{e^2(\kappa-1)\Delta^\kappa}{2^{\kappa-1}\kappa!T^{\kappa-2}}\cos(2\pi\phi_x),
\end{eqnarray}
where $I_i(x)$ are the Bessel functions.

 Let us now turn to
time-dependent contribution to $\Pi (\tau )$. With the aid of Eqs.
(\ref{pi}), (\ref{cspe1}) and (\ref{mcor}) we identify
\begin{equation}
\mathcal P_{osc}+\tilde{\mathcal P}f(\tau)= \frac{1}{\mathcal
Z}\sum\limits_{m>n}|\langle m|\hat I|n \rangle|^2
e^{-\beta\varepsilon_n}e^{-\tau(\varepsilon_m-\varepsilon_n)}.
\label{pf}
\end{equation}
The form of this -- non-universal -- contribution cannot be
recovered within the tight-binding model as it explicitly depends
on the instanton solution and, hence, on the particular shape of
the periodic potential.

In order to proceed let us observe that there exist $\kappa$
low-lying quantum levels with energies
$\Omega/2-\Delta\cos\left(2\pi(\phi_x-k)/\kappa\right)$ in our
problem. These states originate from tunneling depletion of the
ground state energy level $\Omega/2$ in each of $\kappa$ potential
wells. Below we will label these states as $|0k\rangle$ with
$k=1,...,\kappa$. Due to the rotation symmetry of our model all
matrix elements between these states vanish and, hence, do not
contribute to the current-current correlation function.

Next $\kappa$ energy levels $|1l\rangle$ with $l=1,...,\kappa$
occur due to depletion of the first excited state $3\Omega/2$.
These states are characterized by the energies
$3\Omega/2+\tilde\Delta\cos\left(2\pi(\phi_x-l)/\kappa\right)$,
where the parameter $\tilde\Delta$ is to be defined below. With
the aid of the symmetry arguments one can again demonstrate that
the matrix elements of the current operator between the states
$|1l\rangle$ with different $l$ vanish while the matrix elements
between the states $|0k\rangle$ and $|1l\rangle$ remain non-zero
provided $k=l$. In order to evaluate these matrix elements it
suffices to consider the instanton contribution small by setting
$(\Delta+\tilde \Delta)\tau\ll1$ in Eq. (\ref{pf}) and to expand
the right-hand side of this equation in powers of $\Delta \tau$
and $\tilde \Delta\tau$. Comparing the first two terms of this
expansion with $\mathcal P_{osc}$ and $\tilde{\mathcal P}f$
(\ref{tpp})-(\ref{Posc}), in the limit $U_0\gg\Omega$ considered
here we identify
\begin{equation}
 \langle 0k|\hat I|1l
 \rangle|^2\approx\delta_{lk}\frac{e^2\Omega^3}{8\pi^2\kappa^2U_0},
\quad \tilde \Delta\approx\frac{32U_0}{\Omega}\Delta .
\end{equation}
 Note that exactly the same expressions can also be recovered
 from the WKB analysis of the Schr\"odinger equation for the
 cosine potential.

\section { PC noise power}

 Finally let us evaluate the real time
current noise power $S_\omega$ defined in Eqs. (\ref{s}),
(\ref{cspe2}). Employing the above results at $T \ll \Omega$ we
obtain
\begin{eqnarray}
 S_\omega=4\pi\mathcal P\delta(\omega)+\frac{e^2\Omega^3}{4\pi\kappa^2U_0\mathcal Z}
 \sum\limits_{k=1}^\kappa e^{\beta\Delta\cos\left(\frac{2\pi(\phi_x-k)}{\kappa}\right)}
 \nonumber\\ \times\left(\delta(\omega-\Omega-\epsilon_k)+\delta(\omega+\Omega+\epsilon_k)\right)
\label{Sw}
\end{eqnarray} where we defined
\begin{equation}
\epsilon_k=\frac{32U_0\Delta}{\Omega}\cos\left(\frac{2\pi(\phi_x-k)}{\kappa}\right).
\end{equation}
We observe that -- in agreement with our general analysis -- PC
noise power has the form of peaks at frequencies equal to the
distance between the energy levels with non-zero matrix elements
of the current operator plus an additional peak at zero frequency.
In the zero temperature limit $T \to 0$ the amplitude of this peak
tends to zero along with the terms related to transitions to
higher energy levels and Eq. (\ref{Sw}) reduces to
\begin{equation}
 S_\omega=\frac{e^2\Omega^3}{4\pi\kappa^2U_0}\left(\delta(\omega-\Omega-\epsilon_0)+\delta(\omega+\Omega+\epsilon_0)\right),
\label{peak1}
\end{equation}
where $\epsilon_0(\phi_x)=$ max$_k\epsilon_k(\phi_x)$. This result
demonstrates again that PC fluctuations indeed persist down to
$T=0$ in which case peaks of PC noise power $S_\omega$ occur at
frequencies corresponding to transitions between the two lowest
energy levels for which the matrix elements of the current operator differ from zero. We also note that $S_\omega$ differs from zero even at zero external flux $\phi_x=0$ when the average PC value is zero. In the presence of dissipation due to
interaction of the particle with other (quantum) degrees of
freedom the energy levels acquire a finite width, the peaks get
broadened and the noise power should differ from zero also in a
wider range of frequencies.

\begin{figure}
\begin{center}
 \includegraphics[width=3.375in]{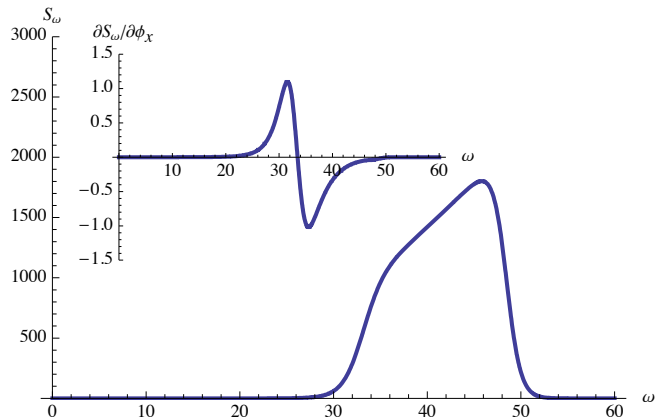}
\end{center}
\label{f2}
\caption{Zero temperature PC noise spectrum $S_\omega$
(arbitrary units) and its derivative with respect to the
    flux $\partial S_\omega/\partial \phi$ (arbitrary units) as
functions of $\omega$ (measured in units of $1/2MR^2$) for an ensemble of
rings (or for a ring containing many independent channels) with $U_0$
uniformly distributed within the interval from $30/MR^2$ to $65/MR^2$.}
\end{figure}

Similarly, broadening of such peaks inevitably occurs in ensembles
  of rings or individual rings with many conducting channels. Within our model
this broadening can be illustrated by considering an ensemble of rings with
the parameter $U_0$ uniformly distributed within some energy interval,
e.g., as it
is indicated in figure caption to Fig. 3. In this case the total PC noise
produced by the system is given by the sum of a large number of very close
peaks (\ref{peak1}) effectively resulting in a much smoother and broader noise
spectrum, as it is shown in Fig. 3.

A specific feature of PC noise is the dependence of $S_\omega$ on the external
magnetic flux $\phi_x$. This dependence occurs due to the presence of quantum
coherence in the system and disappears if this coherence gets destroyed.
Hence, such sensitivity of PC noise spectrum to the flux
can be used as a measure of quantum coherence in our system. Taking the
derivative of $S_\omega$ with respect to the flux, for the model considered
here we obtain
\begin{equation}
\partial S_\omega/\partial \phi_x \propto \sin (2\pi \phi_x/\kappa ).
\end{equation}
Typical dependence of  $\partial S_\omega/\partial \phi_x$ on $\omega$ is
illustrated in Fig. 3. We believe that the main qualitative features of our
results displayed in Fig. 3 should survive also in other models and can be
detected in experiments with nanorings.

It is also interesting to point out a direct physical analogy
between our results and those of Refs. \cite{Averin,Madrid,GZ10} where
equilibrium supercurrent noise in point contacts between
superconductors was investigated. Also in that case the noise
power spectrum has the form of peaks which occur both at zero
frequency and at frequencies equal to the distance between Andreev
levels inside the contact. At $T\to 0$ the zero frequency peak
disappears while the other peaks do not vanish except in the limit
of fully transparent barriers. In the case of many channel
diffusive contacts the supercurrent noise spectrum gets broadened
\cite{GZ10} in a qualitatively similar way to the result displayed
in Fig. 3. In addition, the noise spectrum \cite{Averin,Madrid,GZ10}
turns out to depend on the phase difference across the superconducting
weak link. This dependence has the same physical origin as
the flux dependence of PC noise considered here.

In summary,  we investigated equilibrium fluctuations of
persistent current in nanorings and demonstrated that these
fluctuations do not vanish even at $T=0$ provided the current
operator does not commute with the total Hamiltonian of the
problem. A specific feature of PC noise is its
quantum coherent nature implying that the noise spectrum
can be tuned by an external magnetic flux inside the ring.
We believe that the key features captured by our
analysis will survive also in other models and can be verified in future
experiments.
Our further analysis will be devoted to the effect of
dissipation on PC fluctuations in systems with many degrees of
freedom.

We are grateful to S.M. Apenko and V.V. Losyakov for valuable discussions.
This work was supported in part by RFBR grant 09-02-00886. A.G.S.
also acknowledges support from the Council for grants of the Russian
President Grant No. ÌÊ-89.2009.2, from
the Landau Foundation and from the Dynasty Foundation.

\end{document}